\def\lanl#1{[#1]}

\documentstyle[preprint,aps]{revtex}
\begin{document}
\draft
\title{ \hfill{\normalsize  hep-th/9601004
} \\[-14pt]
\hfill {\normalsize UCSBTH-95-39}\\[-16pt]
\hfill{\normalsize NSF-ITP-95-164} \\
\smallskip
An Extreme Black Hole with Electric Dipole Moment}
\author{Gary T. Horowitz \footnote{email: gary@cosmic.physics.ucsb.edu}\\
and \\
Tsukasa Tada\footnote{email: tada@itp.ucsb.edu}}
\address{Department of Physics \\
University of California \\
Santa Barbara, CA 93106, USA \\
 }

\maketitle
\begin{abstract}
We construct a  new extreme black hole solution in toroidally 
compactified heterotic string
theory.  The black hole saturates the Bogomol'nyi bound, has 
zero angular momentum, but nonzero
electric dipole moment.
It is obtained by starting with a higher
dimensional rotating charged black hole, and compactifying one 
direction in the plane of rotation.
\end{abstract}

\vfill \eject

The black hole uniqueness theorems show that any static black hole
with a regular event horizon must be spherically symmetric. This
implies that all multipole moments vanish except for the mass and charge.
Uniqueness theorems have been proven for black holes in vacuum \cite{ISRR},
Einstein-Maxwell
theory \cite{ISR}, and more recently, Einstein-Maxwell-dilaton theory
\cite{MASOOD}. However, when the dilaton is present, the event horizon
may shrink down to zero size in the extremal limit and become singular.
In this case, the theorems no longer apply. Of course any limit
of static, nonextremal black holes must remain spherically symmetric. 
However, it has recently been realized that extreme black holes are
qualitatively different objects and may have properties which are
not shared by their nonextreme analogs.

In this letter we will construct a static 
extreme black hole with nonzero electric
dipole moment. While the magnitude of the charge is related to the mass
by the extremality condition, the dipole moment is arbitrary, and  is an
independent parameter in the solution. The solution involves more than one
$U(1)$ gauge field, and the dipole moment is associated with
a different gauge field than the charge.

We will work in the context of heterotic string theory compactified on a
torus.  Our starting point is the observation \cite{HORSEN}
that in more than five dimensions, extremally charged rotating  black holes 
saturate a Bogomol'nyi bound. This is in contrast to the situation in
lower dimensions where the extremal limit  involves the angular momentum
and does not saturate this  bound. When the Bogomol'nyi bound is
saturated, there is no force between the black holes and they can
be superposed.  By taking an infinite periodic array of these higher
dimensional black holes, one
can compactify some of the spatial directions and obtain effectively
four dimensional solutions. This approach was used in \cite{HORSEN}
to obtain rotating  black holes in four dimensions which saturate
the Bogomol'nyi bound. To preserve the angular momentum, it was necessary
to compactify directions which were orthogonal to the plane of rotation.
We will show that by compactifying a direction in the plane of rotation,
one obtains a nonrotating extreme black hole with  nonzero dipole moment.

To begin, we review the six-dimensional, extremally charged, rotating black  
hole
solution in  heterotic string theory compactified on a four dimensional  
torus
\cite{HORSEN}. Since the scalar moduli fields remain constant for this
solution\footnote{We consider black holes with ``right moving" charge
only. For a more general solution, see \cite{HORSEN}.}, it suffices to work with the following low energy effective action 
\begin{equation}
\label{e1}
S = 
\int d^6 x \sqrt{-\det G} \, e^{-\Phi} \, \Big[ R_G + 
\partial_\mu \Phi \partial^\mu\Phi -{1\over 12}H_{\mu\nu\rho} H^{\mu\nu\rho}
 - \sum_{j=1}^4 F^{(j)}_{\mu\nu} F^{\mu\nu(j)} \Big]
\end{equation}
where
\begin{equation}\label{e2}
F^{(j)}_{\mu\nu} = \partial_\mu A^{(j)}_\nu - \partial_\nu A^{(j)}_\mu \, ,
\end{equation}
\begin{equation}
\label{e3}
H_{\mu\nu\rho} = \partial_{\mu}B_{\nu\rho} + 2 \sum_{j=1}^4 A_{\mu}^{(j)}
F^{(j)}_{\nu\rho} +\hbox{cyclic permutations of $\mu$, $\nu$, $\rho$}\,.
\end{equation}
$A^{(j)}_\mu$ are four $U(1)$ gauge fields that arise from the compactification
on $T^4$, and $R_G$ is the scalar curvature associated with the metric
$G_{\mu\nu}$.
To describe the solution, it 
is convenient to introduce $\rho$ and $\theta$ which are
related
to the Cartesian coordinates $x^1, \dots, x^5$ and a parameter $a$
(which will be related to the angular momentum) as follows;

\begin{eqnarray}
(x^1)^2+(x^2)^2=(\rho^2+a^2) \sin ^2 \theta , \\
 R^2 \equiv \sum_{i=1}^{5} (x^i)^2 = \rho^2+a^2 \sin^2\theta \, .
\end{eqnarray}
An extremally charged black hole 
which
is rotating in the $ x^1, x^2$ plane can be described in IWP form
\cite{IWP,KALLOSH} in terms of
a function $F(\vec x)$ and vector $\omega_i(\vec x)$ as follows \cite{HORSEN};
\begin{equation}
\label{enx1}
ds^2 = - F^2(\vec x) [dt+ 
\omega_i(\vec x) dx^i]^2
+  d\vec x^2\, ,
\end{equation}
\begin{equation}
\label{enx3}
A^{(j)}_t={p^{(j)}\over \sqrt 2}
 [F - 1]\, , \qquad A^{(j)}_i={p^{(j)}\over \sqrt 2}
 F \omega_i \qquad 
(j=1,\dots,4) ,
\end{equation}
\begin{equation}
\label{enx5}
\Phi =  \ln F(\vec x)\, , \qquad 
B_{ti} = - F \omega_i \, , \qquad B_{ik} =0 \, ,
\end{equation}
where
\begin{equation}
\label{enx6}
F^{-1}  =  1 +   {  m_0 \over \rho(\rho^2 + a^2 \cos^2 \theta)}
\, , \qquad
\omega_i dx^i =
{m_0 a (x^1 dx^2-x^2 dx^1)\over \rho(\rho^2+a^2\cos^2\theta)(\rho^2+a^2)} ,
\end{equation}
and $ p^{(j)}$ is a set of constants which satisfy $\sum_{j=1}^{4}
(p^{(j)})^2=1$.

Several comments on this solution are in order. The mass, total charge, 
and angular 
momentum are given by $M= 3m_0/2, \ Q^2 \equiv \sum_{j=1}^4 (Q^{(j)})^2 =
9m_0^2/ 2$, and $J = m_0 a/2$. So $M^2 = Q^2/2$ as expected for an extreme
dilatonic black hole.
Notice that $J$ is independent of 
$M$ even though the black hole is extremal. 
This is related to the fact that rotating uncharged black holes exist
in six dimensions for all values of $M$ and $J$ \cite{MYPE}. There is no
extremal limit in this case. Another unusual property is that there is no
ergosphere; the Killing vector $\partial/\partial t$ is timelike everywhere.
The surface $\rho = 0$ is a curvature singularity which is null.
This is  a result of the event horizon shrinking down to zero size.
The spacetime is free of naked (timelike) singularities. 
Finally, this solution arises from the dimensional reduction of a
`chiral null model' \cite{HORTSE2} 
which does not receive $ \alpha '$ corrections  
in
a particular renormalization scheme.

For configurations of the form (\ref{enx1} - \ref{enx5}), the field equations
derived from (\ref{e1}) reduce to the following linear equations for 
$F^{-1}$ and $\omega_i$  
\begin{equation}
\sum_{i=1}^{5} \partial_i \partial_i F^{-1} =0, \qquad
\sum_{i=1}^{5}\partial_i \partial_{[i} \omega _{j]}=0,
\end{equation}
One can thus construct multi-black hole solutions by superposing 
these single black  
holes.
This is a reflection of the fact that there is no force between
objects which saturate a Bogomol'nyi bound. In the following we study black
hole solutions obtained by superposing the solution (\ref{enx1}-\ref{enx6})  
and
its spatial translation.

If we place an  infinite number of black holes along a line a
unit distance apart,
the resulting space time has a periodicity one along the line.
This procedure is equivalent to compactifying one dimension on a
torus with the period one\cite{HASH,MYERS,KHURI,GAUHAR}.
By putting  six dimensional
black holes spinning in the $ x^1,\ x^2$ plane along  
the
$ x^4$ and $ x^5$ directions one obtains a four-dimensional rotating black  
hole
which saturates the Bogomol'nyi bound without naked singularities\cite{HORSEN}.

We are interested in compactifying one of the directions in the plane of
rotation e.g. the $ x^1$ direction.
The asymptotic values of $ F^{-1}$ and $ \omega_i$ for a single six dimensional
black hole
are given by\footnote{Expanding to higher order, one finds that $ F^{-1}$
has a 
quadrupole
moment term  ${ m_0
a^2\left(\frac32\sum_{i=1}^{2}(x^i)^2-\sum_{i=3}^{5}(x^i)^2\right)/ R^7}$ 
but no dipole moment.
\label{footnote}}
\begin{equation}
\label{ekk4}
F^{-1}\simeq 1 +{m_0\over R^{3}}\, ,
\qquad
\omega_i dx^i \simeq {m_0 a (x^1 dx^2-x^2 dx^1) \over R^{5}}\, .
\end{equation}
A periodic array of black holes in the $ x^1$ direction with periodicity one
then yields the following asymptotic values for $ F^{-1}$ and $ \omega_i$:
\begin{equation}
\label{ekk5}
 {\breve F}^{-1} \simeq 1 + \sum_{n=-\infty}^\infty {m_0 \over \{  
\breve{R}^2 +
(x^{1}-n)^2\}^{3\over 2}}\, ,
\end{equation}
\begin{equation}
\label{ekk6}
\breve{ \omega}_i dx^i \simeq \sum_{n=-\infty}^\infty {m_0 a \{(x^1-n)dx^2 -
x^2  dx^1\}\over
\{ \breve{ R^2} + (x^{1}-n)^2\}^{5\over 2}}\,
\end{equation}
where
\begin{equation}
\breve{ R}^2 \equiv  \sum_{i=2}^{5} (x^i)^2 = R^2-(x^1)^2.
\end{equation}
For $ {\breve R}\gg 1 $ the summations on the right hand side of (\ref{ekk5})
and (\ref{ekk6}) can be evaluated by integration as follows\footnote{Similar
sums can be evaluated analytically with the result that the exact
answer differs from the leading term computed here only by exponentially
small $ O(\exp(-{\breve R}))$ contributions. (See  e.g. \cite{HASH}.)}
\begin{equation}
{\breve F}^{-1} \simeq  1 + \int_{-\infty}^\infty du {m_0 \over \{  
\breve{R}^2
+
(x^{1}-u)^2\}^{3\over 2}}\, ,
\end{equation}
\begin{equation}
\breve{ \omega}_1  \simeq -\int_{-\infty}^\infty du{m_0 a x^2 \over
\{ \breve{ R^2} + (x^{1}-u)^2\}^{5\over 2}}\, \ \  ,\qquad \breve{ \omega}_2
\simeq  \int_{-\infty}^\infty du {m_0 a (x^1-u)   \over
\{ \breve{ R^2} + (x^{1}-u)^2\}^{5\over 2}}\, .
\end{equation}
By changing variable $ u \equiv x^1 + \breve{R} v$ one can easily perform  
the
integrals above and obtain
\begin{equation}
{\breve F}^{-1} \simeq 1 + {m_0 \over \breve{R}^2}\int_{-\infty}^\infty dv   
{1
\over (1+v^2)^{{3\over 2}}}= 1 +{2 m_0 \over \breve{R}^2},
\end{equation}
\begin{equation}
\breve{ \omega}_1  \simeq -{m_0 a \over \breve{R}^4}x^2 \int_{-\infty}^\infty  
dv
 {1 \over (1+v^2)^{{5\over 2}}}=-{4 m_0 a \over 3 \breve{R}^4} x^2 \,\, ,
\end{equation}
\begin{equation}
\breve{\omega}_2 \simeq -{m_0 a \over \breve{R}^3}  
\int_{-\infty}^\infty
dv  {v \over (1+v^2)^{{5\over 2}}}=0 .
\end{equation}

To proceed, there are essentially two choices for the further  
compactification
direction. If we compactify the second direction in the rotation plane,
$ x^2$, with the same
procedure we find
\begin{equation}\label{ChargeF}
\hat{F}^{-1} \equiv 1+\sum_{n=-\infty}^{\infty}
[\breve{F}^{-1}(t,x^1,x^2-n,x^3,x^4,x^5)-1] \simeq
1+ {2\pi m_0 \over \hat{R}} \, ,
\end{equation}
\begin{equation}\label{Chargeo}
\hat{\omega}_{i} \equiv \sum_{n=-\infty}^{\infty}
\breve{\omega}_{i}(t,x^1,x^2-n,x^3,x^4,x^5) \simeq 0 \, ,
\end{equation}
where
\begin{equation}
\hat{ R}^2 \equiv  \sum_{i=3}^{5} (x^i)^2 .
\end{equation}

In this case, the asymptotic geometry is that of the standard
extremally charged  
dilatonic
black hole \cite{GIMA,GHS}
\begin{equation} \label{cbh}
ds^2=-\left( 1+ {2\pi m_0 \over \hat{R}} \right) ^{-2} dt^2 +
\sum_{i=3}^5(dx^i)^2\, ,
\end{equation}
\begin{equation}
\Phi=- \ln \left( 1+ {2\pi m_0 \over \hat{R}} \right) \,  ,
\end{equation}
\begin{equation}
A^{(j)}_t={p^{(j)}\over \sqrt 2}
 \left[ \left( 1+ {2\pi m_0 \over \hat{R}} \right)^{-1} - 1\right]   \,  
,
\end{equation}
\begin{equation}\label{cbhh} A^{(j)}_i=0\qquad    \qquad (j=1,\dots,4 )\, .
\end{equation}
All other fields vanish. This includes
the four-dimensional components of the anti-symmetric tensor $
B_{\mu\nu}$, and additional $U(1)$ gauge fields and scalars that could
arise in the
compactification down to four dimensions.
The above  expressions for the geometry, $U(1)$ gauge fields and scalar dilaton
agree with the standard four-dimensional charged dilatonic black hole 
with mass
$M= \pi m_0$ and total charge $Q^2= 2 \pi^2
m_0^2$. Although we derived (\ref{cbh}-\ref{cbhh}) keeping only the
leading order terms for large $\hat{R}$, it nevertheless accurately
describes the four dimensional solution down to the compactification scale.
This is because the original solution treated the coordinates
$x^i, \ i = 3,4,5$ symmetrically, so higher order terms in the expansion of
$F$ and $\omega_i$ cannot lead to higher order multipole moments.

The more interesting alternative is to choose our second direction to lie
orthogonal to the rotation plane e.g. $ x^5$.
We then find
\begin{equation}\label{DipoleF}
\bar{F}^{-1} \equiv 1+\sum_{n=-\infty}^{\infty}
[\breve{F}^{-1}(t,x^1,x^2,x^3,x^4,x^5-n)-1] \simeq
1+ {2\pi m_0 \over \bar{R}} \, ,
\end{equation}
\begin{equation}\label{Dipoleo}
\bar{\omega}_{1} \equiv \sum_{n=-\infty}^{\infty}
\breve{\omega}_{1}(t,x^1,x^2,x^3,x^4,x^5-n) \simeq - {2 \pi m_0 a\over 3
\bar{R}^3} x^2\, , \qquad\bar{\omega}_2 \simeq 0 \, ,
\end{equation}
where
\begin{equation}
\bar{R}^2\equiv \sum_{i=2}^{4} (x^{i})^2 \, .
\end{equation}
Asymptotically, the resulting geometry  is given by:
\begin{eqnarray}\label{dipolegeo}
ds^2 &&= -\left( 1+ {2\pi m_0 \over \bar{R}} \right) ^{-2} [dt- {2 \pi m_0
a\over 3 \bar{R}^3}  x^2dx^1]^2
+   \sum_{i=1}^4(dx^i)^2 \, \nonumber \\
&&=-\left( 1+ {2\pi m_0 \over \bar{R}} \right) ^{-2} dt^2+{4 \pi m_0 a
x^2\over 3 \bar{R}^3\left( 1+ {2\pi m_0 \over \bar{R}} \right) ^{2}}  dt  
dx^1
\nonumber \\
&&\qquad  + \left( 1- {4 \pi^2 m_0 ^2 a ^2 (x^2)^2\over 9 \bar{R}^6\left( 1+
{2\pi m_0 \over \bar{R}} \right) ^{2}}\right) (dx^1)^2 +\sum_{i=2}^4  
(dx^i)^2.
\end{eqnarray}
Since $x^1$ is now a compact direction, we can  extract the four dimensional
fields from this five dimensional metric using the 
Kaluza-Klein decomposition
\begin{equation}
ds^2=[g_{\mu\nu}(x)+A_\mu(x) A_\nu(x)] dx^\mu dx^\nu+2 A_\mu dx^\mu dy+
e^{2\sigma}dy^2 \, ,
\end{equation}
where $\mu = 0,2,3,4$ and $y=x^1$. The result is
\begin{eqnarray}
e^{2\sigma} &=&1-{4 \pi^2 m_0 ^2 a^2 (x^2)^2\over 9 \bar{R}^6
\left( 1+ {2\pi m_0 \over
\bar{R}} \right) ^{2}} \, \simeq 1 \, \  ,\label{bransdicke} \\ \nonumber \\
A_t&=&{2 \pi m_0 a x^2\over 3 \bar{R}^3\left( 1+ {2\pi m_0 \over \bar{R}}
\right) ^{2}} \, \simeq \, {2 \pi m_0 a \over 3 \bar{R}^3}x^2\ ,\label{At}  
\\
\nonumber \\
A_i&=& 0   \qquad \hbox{for}  \qquad i=2,3,4
\end{eqnarray}
\begin{eqnarray}\label{mfd}
g_{tt}&=&-\left( 1+ {2\pi m_0 \over \bar{R}} \right) ^{-2}\left( 1+ {4 \pi^2
m_0^2 a^2  (x^2)^2\over 9 \bar{R}^6\left( 1+ {2\pi m_0 \over \bar{R}}  
\right)
^{2}} \right)
\, \simeq -\left( 1+ {2\pi m_0 \over \bar{R}} \right) ^{-2}\, ,\label{gtt}  
\\
\nonumber \\
g_{ti}&=&0, \qquad g_{ik} = \delta_{ik} \qquad \hbox{for} 
\qquad i,k=2,3,4 \,.
\end{eqnarray}
where we have kept only the leading order terms on the right. Notice that
the angular momentum is zero, but the
gauge field $A_\mu$ clearly has a nonzero electric dipole moment which is
proportional to $a$, and pointing in the $x^2$
direction. Reducing the antisymmetric tensor field yields another $U(1)$ gauge 
field
in four dimensions $B_{\mu 1}$ with exactly the same dipole moment
\begin{eqnarray}
B_{t1}&=&-\bar{F}\bar{\omega}_1\simeq {2 \pi m_0 a \over 3
\bar{R}^3}x^2\ ,\label{Bt1}  \\
B_{i1}&=&0    \qquad \hbox{for}\quad i=2,3,4\, .\label{Bi1}
\end{eqnarray}
The four dimensional antisymmetric tensor field vanishes 
$B_{ti}=B_{ik}=0$.
Finally, the gauge fields that were already present in six dimensions
reduce to give four dimensional gauge fields
\begin{equation}\label{gffd}
A^{(j)}_t={p^{(j)}\over \sqrt 2}
 \left[\left( 1+ {2\pi m_0 \over \bar{R}} \right)^{-1} - 1\right]   \,  
,
\end{equation}
\begin{equation} A^{(j)}_i=0\qquad   \hbox{ for} \quad i=2,3, 4\qquad
(j=1,\dots,4 )\,  ,
\end{equation}
as well as  nonzero scalars
\begin{equation}
A^{(j)}_1={p^{(j)}\over \sqrt 2}
 \bar{F} \bar{\omega}_1  
\simeq -{\sqrt2 \pi m_0 a \over 3 \bar{R}^3}p^{(j)}x^2\qquad(j=1,\dots,4)\, .
\end{equation}
{}From (\ref{mfd}) and (\ref{gffd}) it is clear that the mass and charge of this
solution is independent of the value of the electric dipole moment.
Therefore in four dimensions the resulting geometry is that of the  
extremally
charged dilatonic black hole with an additional dipole moment in 
the $ x^2$ direction. Higher order terms in the expansion of $F$ and
$\omega_i$ will now yield higher order multipole moments as well, but
their values are determined in terms of $m_0$ and $a$.

There is increasing evidence that massive BPS string states
are  described by extreme black holes.
Are there elementary string states which correspond 
to these black holes with electric dipole moment? If the dipole moment is not
too large, the answer is yes. This follows from \cite{DGHW} where
the fields outside of
certain macroscopic string sources wrapped around a compact direction in
$D$ dimensions were constructed and shown to 
reduce to rotating black holes in $D-1$ dimensions. (These solutions were
also discussed in \cite{CMP}.) Consider a string in
six dimensions that reduces to a black hole rotating in the
$x^1,\ x^2$ plane in five dimensions. Since these fields can
also be superposed, one can compactify the $x^1$ direction by taking 
an infinite chain
of such strings.   One can show that the
resulting four dimensional solution again describes
a nonrotating black hole with
electric dipole moment. Since the angular momentum associated with
the string source satisfies the Regge bound, the parameter $a$ is
bounded, and the dipole moment cannot
take arbitrarily large values.

To summarize, we have studied extreme black holes in heterotic string
theory which are rotating in a compactified direction. From the four
dimensional viewpoint, these solutions describe nonrotating black
holes with  electric dipole moments. Although we have discussed only
electric charges and dipole moments, one can clearly perform an $S$-duality
transformation on the four dimensional fields to obtain magnetic 
charges and dipole moments. Are more general solutions possible
with independent higher multipole moments? (See \cite{RIW} for an 
interesting example of nonspherical black holes in a different context.)
Since the extreme black hole
has a singularity at the origin and no real event horizon, it might appear
easy to construct a large class of similar static solutions with 
$M^2 = Q^2/2$ and arbitrary multipole moments. However, in general
these solutions will have timelike singularities at the origin rather
than the more mild null singularities found here. Furthermore, they are
unlikely to be supersymmetric. The solution constructed here has unbroken
supersymmetry (at least to leading order in $\alpha'$) since it arises 
from the dimensional reduction of a chiral null model \cite{HORTSE2}.

It has been shown that black holes with both electric and magnetic charges
can have nonzero horizon area even in the extremal limit \cite{KL,CY}.
It would be
interesting to construct the rotating analog of these solutions in higher
dimensions and
compactify one direction in the plane of rotation. The result would appear
to be a nonrotating black hole with dipole moment and nonzero 
horizon area. This seems to contradict the uniqueness theorems.
(Although a theorem has not yet been rigorously established for
this case, it is widely believed to hold.) The resolution of this apparent
contradiction will certainly deepen our understanding of the properties of
black holes.

{\bf Acknowledgment}: GTH was supported
in part by NSF Grant PHY95-07065. TT is partially supported by NSF Grant
PHY89-04035 and the JSPS Postdoctral Fellowship for Research Abroad.

\end{document}